\documentclass[prd,nofootinbib]{revtex4}

\usepackage{epsfig}
\usepackage{graphicx}
\usepackage{amssymb,amsmath}
\usepackage{epstopdf}

\newcommand{\be}{\begin{equation}}
\newcommand{\ee}{\end{equation}}
\newcommand{\bea}{\begin{eqnarray}}
\newcommand{\eea}{\end{eqnarray}}

\newcommand\Deighta{{$\overline {\textrm{D}}8$}}

\begin{document}

\title{Holographic Magnetic Phase Transition}

\author{Gilad Lifschytz}
\affiliation{Department of Mathematics and Physics and CCMSC\\
University of Haifa at Oranim\\
Tivon 36006, Israel}
\email{giladl@research.haifa.ac.il}

\author{Matthew Lippert}
\affiliation{Department of Physics\\
Technion, Haifa 32000, Israel\\
{\rm and}\\
Department of Mathematics and Physics \\
University of Haifa at Oranim\\
Tivon 36006, Israel}
\email{matthewslippert@gmail.com}


\begin{abstract}
We study four-dimensional interacting fermions in a strong magnetic field, using the holographic Sakai-Sugimoto model of intersecting D4 and D8 branes in the deconfined, chiral-symmetric parallel phase.  We find that as the magnetic field is varied, while staying in the parallel phase, the fermions exhibit a first-order phase transition in which their magnetization jumps discontinuously.  Properties of this transition are consistent with a picture in which some of  the fermions jump to the lowest Landau level.  Similarities to known magnetic phase transitions are discussed.
\end{abstract}

\maketitle

\section{Introduction}
The study of QCD-like theories using a holographic approach has been a very fruitful line of investigation in the last few years.  Many qualitative features of QCD appear to be reproduced in these holographic models.  In particular, the Sakai-Sugimoto (SS) model \cite{ss1, ss2}, based on the dynamics of D8-branes and \Deighta-branes in the background of near-horizon D4 branes, has yielded many interesting results.  Most studies of the phase diagram \cite{asy, ps, bll1, ucla, stony, bll2, ubc, usc} have focused on the confined phase or on the interplay between the chiral-broken phase and chiral-symmetric phase in the deconfined phase as a function of external parameters such as temperature, density, and electromagnetic fields. 

In this paper we will study relativistic interacting fermions at high density and strong magnetic field using this holographic description. This is done in the Sakai-Sugimoto model in the deconfined and chiral-symmetric phase. In this phase the low-energy excitations are charged fermions. 

As shown in \cite{bll3, ll1}, when a magnetic field is turned on, the system behaves as if the charge is made out of two components, one which is represented holographically in the bulk by sources located at the horizon and the other by smeared D4-branes inside the D8-brane along the radial direction. The first behaves under an applied electric field as charges in a dissipative neutral medium and the second behaves as if no dissipation is present. In this letter we show that, for a given temperature and density, there is a first-order phase transition as a function of the magnetic field for which there is a jump of the charges of the first kind to charges of the second kind along with a jump in the magnetization.  A similar phenomenon in real materials where the magnetization grows rapidly is known as metamagnetism. The properties of the transition seem to indicate that it is consistent with a picture in which a phase transition of the fermions jumping to the lowest landau level has occurred. The  fermions in the lowest landau level seem to form baryonic-like bound states which are singlets under the $SU(N)$ and thus do not interact with the dissipative bath of gluons.  

\section{The model background}

We will consider the SS D4-D8 system in the parallel phase. The D4-D8 intersection has $ND=6$ 
and so has as its lowest excitations only chiral fermions in four dimensions. 
In the phase where the D8-branes and $D\bar{8}$-brane are parallel, chiral symmetry is unbroken 
and we have both left- and  right-handed chiral fermions in the fundamental of $SU(N)$. 
The fermions are charged under the U(1) symmetry on the D8-branes, and this interaction resembles electromagnetism but without dynamical photons. The system has two free parameters, $L$ which is the distance between the $D8$-brane and the $D\bar{8}$-brane, and the size 
$R_{4}$ of the circle in the $x_{4}$ direction. The system goes through various phases
 as a function of these parameters and external conditions such as temperature, 
density, and magnetic field. At low temperature the system is in the confined 
phase and at higher temperature $T>\frac{1}{2\pi R_{4}}$ it is in the deconfined phase.
In the deconfined phase at high enough temperature (which depends on $L$) the system
 will be in the chiral-symmetric phase. The phase transition we will find occurs at any
 temperature and $L$ provided the density and magnetic fields are large enough.

The metric and background fields of the near horizon limit of the D4-branes is given by
\bea
\label{deconfined_background}
ds^2 &=& u^{3\over 2} \left(f(u)(dx_0^E)^2 + d{\bf x}^2
+ dx_4^2\right)
+ u^{-{3\over 2}}\left({du^2\over f(u)}
+ u^2 d\Omega_4^2\right) \ , \nonumber\\
e^{\Phi} &=& g_s u^{3/4} \; , \;
F_4 = 3\pi (\alpha')^{3/2}  N_c \, d\Omega_4 
\eea
where
\be
f(u) = 1 - {u_T^3\over u^3}
\ee
and $u=\frac{U}{R}$,  $x_{\mu}=\frac{X_{\mu}}{R}$, and  $R=(\pi g_{s}N_{c})^{1/3}\sqrt{\alpha'}$.  $u_T$ is related to the temperature, {\em i.e.} the inverse periodicity of the Euclidean time $x_0^E$, as $u_T=(4\pi/3)^2 R^2 T^2$.
In the holographic description the free energy is the action of the D8 branes in the near-horizon limit of the D4-branes. 

Baryonic charge is identified with the charge under the $U(1)$ living on the D8-branes.  Adding charges to the system is accomplished by  sourcing the $A_{0}$ component of the $U(1)$ gauge field.  One can turn on a magnetic field in this $U(1)$ as $2\pi \alpha^{'}F_{23}=h$.  At zero magnetic field the charges sit at the horizon, but at non-zero magnetic field, due to the Chern-Simons term in the world volume action of the D8-brane which holographically encodes the axial anomaly, $A_{1}$ is turned on and induces a baryon charge proportional to $h$ along the D8-brane \cite{bll3}.  In addition, an anomaly-driven axial current $j_{A}=\frac{3}{2}h\mu$ is generated in the direction of the magnetic field.

Generally, when a magnetic field is turned on, for a fixed $L$, there is a possibility that the system will favor a chiral-broken phase. However, we can change $L$ at will and, for a given charge density (or chemical potential), delay breaking of chiral symmetry such that the magnetic phase transition occurs before the chiral phase transition. Whether this is true in QCD is hard to say.

\section{Four-dimensional holographic fermions}

The action of the D8 brane in this background includes both a Dirac-Born-Infeld (DBI)  term and a Chern-Simons (CS) 
term which embodies the anomaly. Through the CS term the magnetic field couples the $A_{0}$ and the $A_{1}$ fields living on the D8 brane. We work with rescaled variables 
\be
a_{\mu}=\frac{2\pi \alpha'}{R}A_{\mu} \; , \; h=2\pi \alpha' F_{23}
\ee
resulting in the action \cite{bll3}
\bea
\label{action}
S_{DBI} &=&   {\cal{N}} \int_{u_T}^\infty du \, u^{5/2} \sqrt{\Big(1
-(a_0'(u))^2 + f(u)(a_1'(u))^2 \Big)\left(1+{h^2\over u^3}\right)} \nonumber \\[5pt]
S_{CS} & = &  - \frac{3}{2}{\cal N}
\int_{u_T}^\infty \left(h a_0(u) a^{'}_{1}(u) 
-h a^{'}_{0}(u) a_1(u)\right), 
\eea
where we have included both the 8-brane and anti-8-brane parts,
with $\bar{a}_0 = a_0$ and $\bar{a}_1=-a_1$, so
\be
{\cal N}=\frac{N_c}{6\pi^{2}}\frac{R^2}{(2 \pi \alpha')^{3}} \ .
\ee

The integrated equations of motion are then given by
\bea
\label{EOMa0dec}
\frac{\sqrt{u^5+h^2 u^2}\, a_0'(u)}{\sqrt{1 - (a_0'(u))^2 + f(u)(a_1'(u))^2}}
&=& 3ha_1(u) + d \nonumber\\
\label{EOMa1dec}
\frac{\sqrt{u^5+h^2 u^2}\, f(u) a_1'(u)}{\sqrt{1 - (a_0'(u))^2 + f(u)(a_1'(u))^2}}
&=& 3ha_0(u) 
\eea
where $d$ is the total rescaled charge density.\footnote{The baryon charge $d$ is related to the physical density D by $D=\frac{2 \pi \alpha' {\cal N}}{R}d$.  In these conventions, quarks (or, more generally, fermions) carry baryon charge $1$ rather than $1/N_c$.} The boundary conditions are $a_{0}(u_{T})=0$ and $a_{1}(\infty)=0$, and there is no constant appearing in the equation of motion for $a_{1}$ due to regularity at the horizon. 

At non-zero magnetic field, some of the charge is found above the horizon and is represented by D4-branes smeared inside the D8-brane.  Like ordinary baryons which holographically correspond to D4-branes localized on the D8-brane, these smeared D4-branes have $N_c$ strings stretching between them and the D8-branes and so correspond holographically to bounds states of $N_c$ quarks.  The amount of charge located above a given $u$ is given by $-3ha_1(u)$, and the total charge in the form of smeared D4-branes $d_*$ is then just $d_* = -3ha_1(u_T)$.  The rest, $d-d_* = d+3ha_1(u_T)$, remains at the horizon.

After a little algebra, the equations of motion (\ref{EOMa0dec}) can be put in the form
\bea
\label{EOMa0dec1}
a_{0}^{'2} &=& \frac{f (3ha_1 + d)^2}{fu^5+fh^2u^2+f (3ha_1 + d)^2-(3ha_0)^2} \nonumber\\
fa_{1}^{'2} &=& \frac{(3ha_0)^2}{fu^5+fh^2u^2+f (3ha_1 + d)^2-(3ha_0)^2} \ . 
\eea

\subsection{Zero temperature}

At any non-zero temperature these equations can only be solved numerically. However, at zero temperature $f=1$, so one can actually solve these equations analytically\footnote{Note that at zero temperature one could in principle add a constant of integration to the once integrated $a_{1}$ equation of motion which represents a current in the system, but this is not relevant to our case.} up to a transcendental equation which is solved numerically. Of course, at zero temperature the chiral-symmetric phase is never the preferred phase \cite{ahjk}, but solving these equations in this case provides more control over the numerics and clarifies the results. As we will see, the numerical solution at low temperature is very close to the zero-temperature results.

To solve the zero-temperature equations, one notices that dividing the two equations in (\ref{EOMa0dec1}) gives
\be
\frac{a_{0}^{'}}{a_{1}^{'}}=\frac{3ha_{1}+d}{3ha_{0}}
\ee
which can be integrated using the boundary conditions $a_{1}(\infty)=0$ and $a_{0}(\infty)=\mu$.  One then finds that
\be
(3ha_1 + d)^2-(3ha_0)^2=d^2-9h^2 \mu^2 \ .
\ee

By defining a new coordinate
\be
z=\int_{0}^{u} \frac{3h \ du}{\sqrt{u^5 +h^2 u^2 + d^2-9h^2 \mu^2}}
\ee
where $z_{\infty}=z(u=\infty)$, the equations (\ref{EOMa0dec1}) reduce to
\bea
\partial_{z} a_{0} &=& a_{1} +\frac{d}{3h} \nonumber \\
\partial_{z} a_{1} &=& a_{0} \ .
\eea
The solution to the equations of motion with the conditions $a_{0}(z=0)=0$ and $a_{1}(z=z_{\infty})=0$ is then
\bea
a_{0} &=& \frac{d \sinh z}{3h\cosh z_{\infty}} \nonumber \\
a_{1} &=& \frac{d \cosh z}{3h\cosh z_{\infty}}-\frac{d}{3h} \ .
\eea
From these equations we see that
\bea
\mu &=& \frac{d}{3h}\tanh z_{\infty} \nonumber \\
a_{1}(0) &=& \frac{d}{3h} \left(\frac{1}{\cosh z_{\infty}} -1\right) \ ,
\eea
and thus,
\be
\label{zintegralequation}
z_\infty=\int_{0}^{\infty} \frac{3h \ du}{\sqrt{u^5 +h^2 u^2 + \frac{d^{2}}{\cosh^{2}z_{\infty}}}} \ .
\ee

\begin{figure}
\centerline{\epsfig{file= 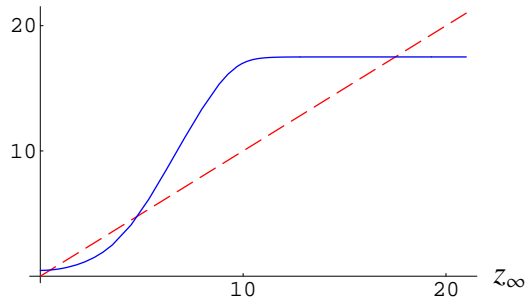,width=7cm}}
\caption{A plot showing the three solutions to the integral equation (\ref{zintegralequation}) for $z_{\infty}$ with $d=1$ and regulator $\epsilon = 10^{-3}$.  The dashed red curve is $z_{\infty}$ while the solid blue curve is the regulated integral on the right hand side of (\ref{zintegralequation}).  The values of $z_{\infty}$ at which the curves cross are solutions.  The regulator is responsible for the plateau in the integral; as $\epsilon$ is decreased, the plateau is pushed higher, and when $\epsilon \to 0$ the largest solution is $z_{\infty} = \infty$.}
\label{zinfinity_fig}
\end{figure}

To find the actual solution for $z_{\infty}(h,d)$, one needs to solve (\ref{zintegralequation}) numerically.  We regularize the integral\footnote{Having to insert by hand a regulator is a particular feature of $T=0$.  At non-zero temperature, $u_T$ naturally cuts off integrals over $u$.} by taking the lower limit of the integral defining $z$ to be $\epsilon$ and at the end sending $\epsilon \rightarrow 0$. 
For small $d$ and $h$ there are three solution for $z_{\infty}$, but for large enough $h$ only one remains which is $z_{\infty} \sim \log \epsilon  \rightarrow \infty$. Fig.~\ref{zinfinity_fig} illustrates the case where there are three solutions.

It should be noted that the appearance of three solution and the interesting phase structure which they imply is due to the use of the full DBI action (\ref{action}) rather than just the Yang-Mills approximation.

\begin{figure}
\begin{tabular}{cc}
\epsfig{file= 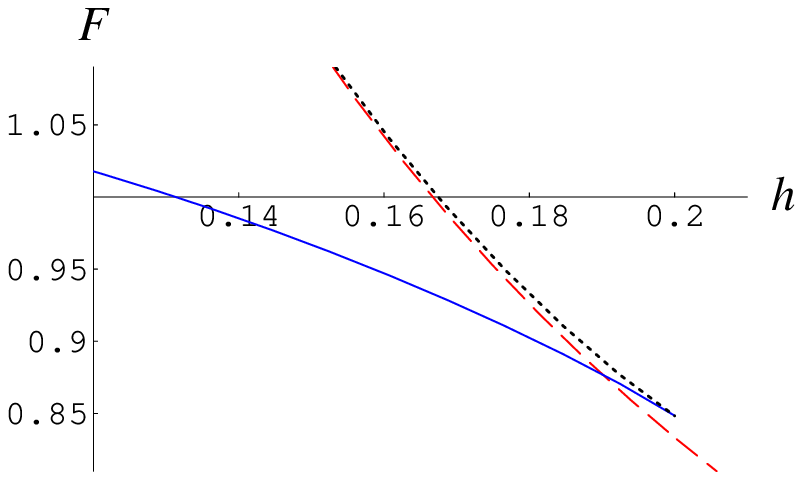,width=7cm} &
\epsfig{file= 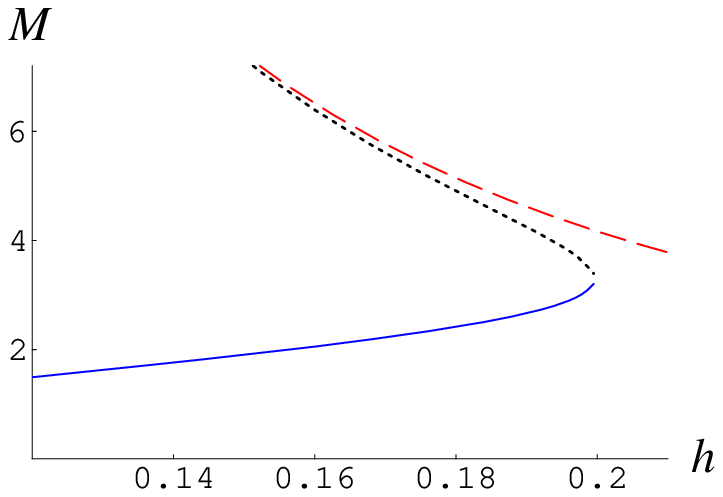,width=7cm}
\end{tabular}
\caption{Three solutions for (a) the free energy $F$ and (b) the magnetization $M$ at $T=0$ with $d =1$.  There is a first-order phase transition from the solid blue $z_\infty \sim h$ solution to the dashed red $z_\infty \to \infty$ solution at $h = 0.19$.  The dotted grey solution, which corresponds to the middle solution in Fig.~\ref{zinfinity_fig}, connects the two stable solutions and is an unstable maximum of the free energy.}
\label{F_vs_h_zeroT_fig}
\end{figure}

The free energy of the three solutions is shown in Fig.~\ref{F_vs_h_zeroT_fig}.  At small $h$ the solution $z_{\infty}\sim h$ is preferred, but at some finite $h \sim 0.19$ there is a jump to the branch parametrized by $z_{\infty} \rightarrow \infty$.  This first-order phase transition is accompanied by jumps in both the magnetization and the chemical potential.

As we can see, the preferred solution at large $h$ has
\be
\label{muandF}
\mu=\frac{d}{3h} \ \ ,\ \ F=\frac{d^2}{6h} \ .
\ee
We also find that $a_{0} = 0$ and $a_{1} = -\frac{d}{3h}$ are constant all the way up to $u \rightarrow \infty$ where they jump to their asymptotic values  $a_{0} = \mu$ and $a_{1} = 0$.  Not only is all the charge smeared inside the D8-branes, $d_* = d$, but it is in fact located right near the boundary at $u=\infty$. 

\subsection{Non-zero temperature}

\begin{figure}[htbp]
\begin{center}
\begin{tabular}{cc}
\epsfig{file= 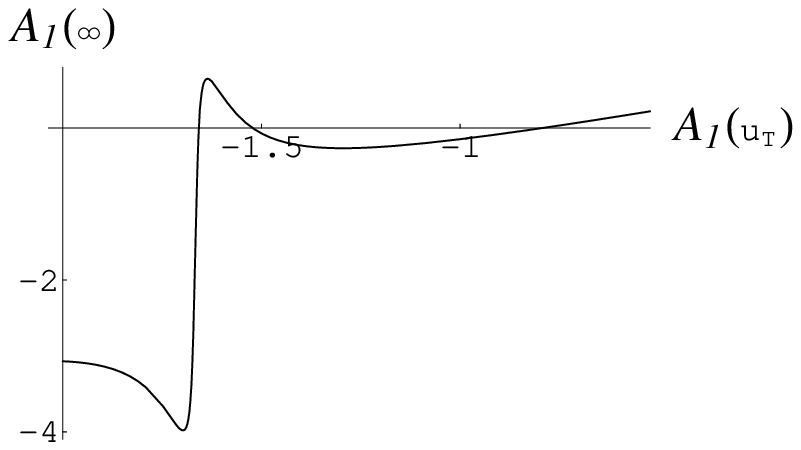,width=7cm} &
\epsfig{file=  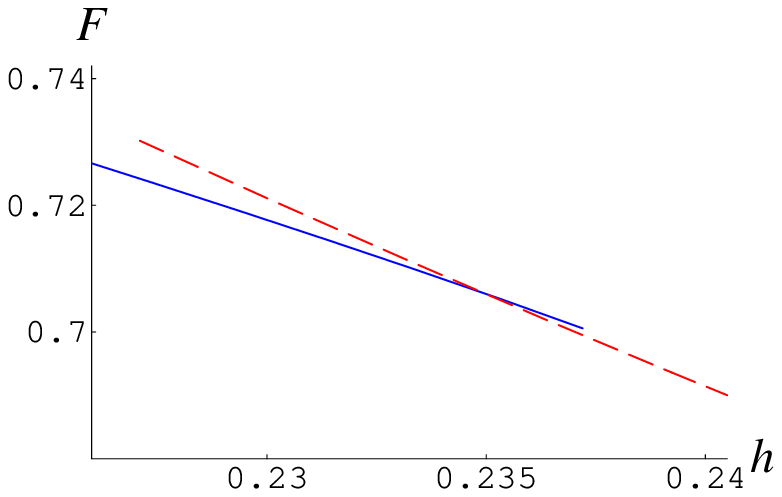,width=7cm}
\end{tabular}
\caption{(a) Three solutions for $A_1(\infty)$ as a function of $A_1(u_T)$ for $n=1$, $T=0.07$, and $h =0.2$.  (b) The free energy $F$ as a function of $h$ for $n=1$ and $T=0.09$.  There are two stable branches, the solid blue curve and the dashed red curve. The phase transition, where the blue and red curves cross, is at $h=0.235$.}
\label{F_vs_h_fig}
\end{center}
\end{figure}

We now turn to solving these equations in the more physical case of non-zero temperature. We will see that the properties seen for zero temperature persist. At high temperature there is only one solution, which means a single phase. As the temperature is lowered, there is a critical temperature $T_{c}(d)$ below which one finds three solutions to the equations of motion for some range of $h$.  One way to see this in the numerics is to plot $a_{1}(\infty)$ as a function of $a_{1}(u_{T})$.  At high temperature the graph (at all values of $h$) cross the horizontal axis only once.  However, at lower temperature there is a range of $h$ when there are three possible values of $a_{1}(u_{T})$ which gives $a_{1}(\infty)=0$. This is shown in Fig.~\ref{F_vs_h_fig}(a).

\begin{figure}[htbp]
\begin{center}
\begin{tabular}{cc}
\epsfig{file= 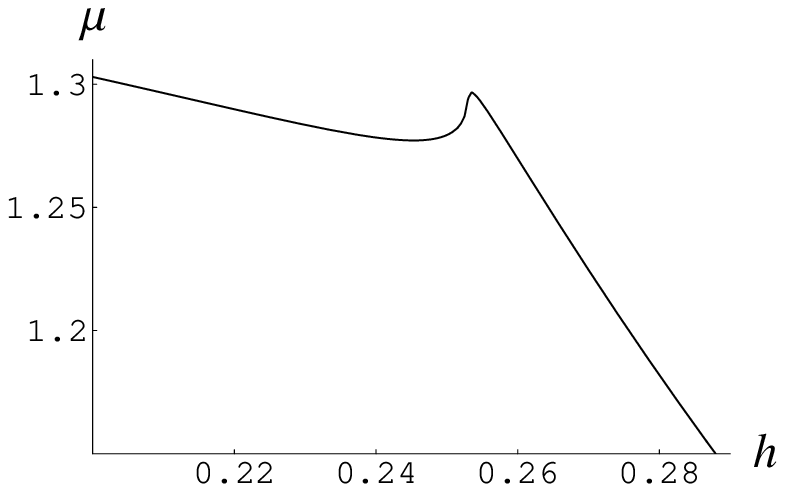,width=7cm} &
\epsfig{file= 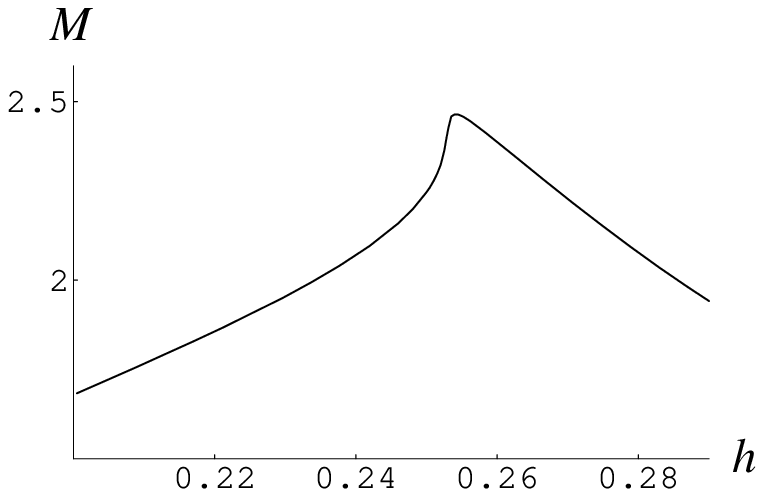,width=7cm}
\end{tabular}
\caption{(a) $\mu$ and (b) $M$ are smooth functions of $h$ for $d=1$ and $T=0.102$ which is just above the critical point at $T=0.101$.}
\label{mu_and_M_vs_h_smooth_fig}
\end{center}
\end{figure}

\begin{figure}[htbp]
\begin{center}
\begin{tabular}{cc}
\epsfig{file= 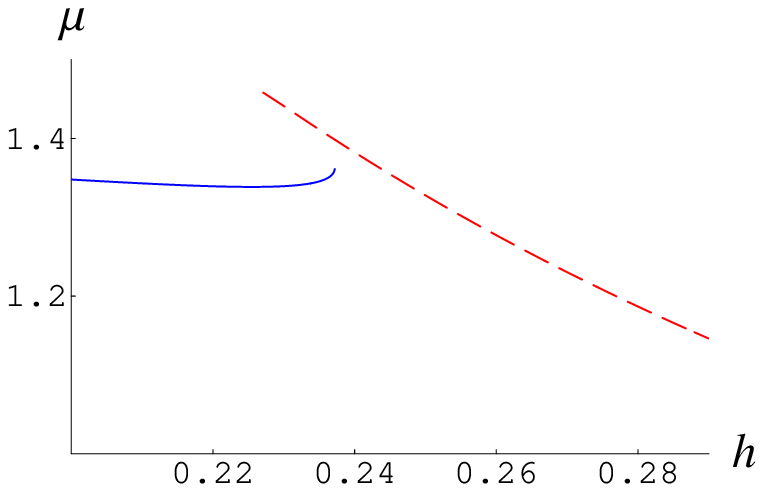,width=7cm} &
\epsfig{file= 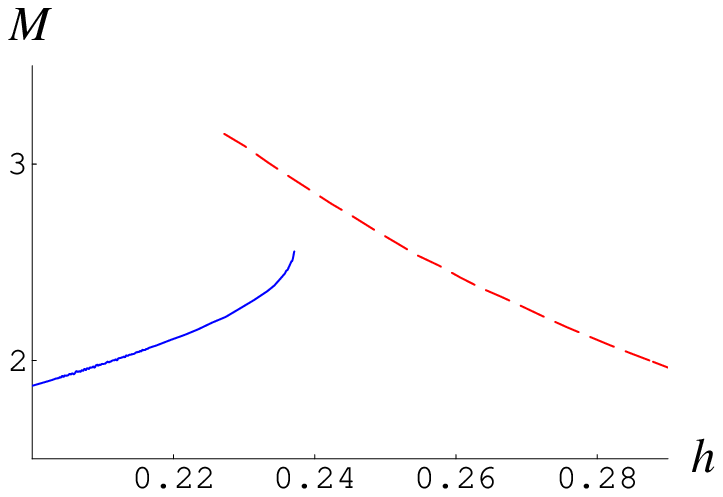,width=7cm}
\end{tabular}
\caption{(a) $\mu$ and (b) $M$ as functions of $h$ for $d=1$ and $T=0.09$, below the critical point.  There are now two branches of stable solutions, and the phase transition between them occurs at $h = 0.235$.}
\label{mu_and_M_vs_h_jump_fig}
\end{center}
\end{figure}

Again, one of the solution is unstable and two are stable. The free energy of the two stable solutions as a function of $h$ can be seen in Fig.~\ref{F_vs_h_fig}(b). We see that there is a phase transition at some non-zero value of $h$. At a temperature slightly higher then the critical temperature  both $\mu(h)$ and $M(h)$ develop what looks like a  wiggle; see Fig.~\ref{mu_and_M_vs_h_smooth_fig}.  As the temperature is lowered, we enter a regime where there is a phase transition; the chemical potential and the magnetization become discontinuous, as shown in Fig.~(\ref{mu_and_M_vs_h_jump_fig}).  At zero temperature, the jump of the chemical potential is such that the axial current $j_{A}=\frac{3}{2}h\mu$ attains its maximum value of $d/2$ and does not change any more as $h$ is increased.  This implies that in the large $h$ phase the system is fully polarized.  At non-zero temperature, $j_{A}$ jumps close to the maximum value and asymptotes to it at large $h$.  Similarly if we had turned on an axial chemical potential rather then a vector chemical potential, there would be a jump in the vector current in the direction of the magnetic field.

\begin{figure}[htbp]
\begin{center}
\begin{tabular}{cc}
\epsfig{file=  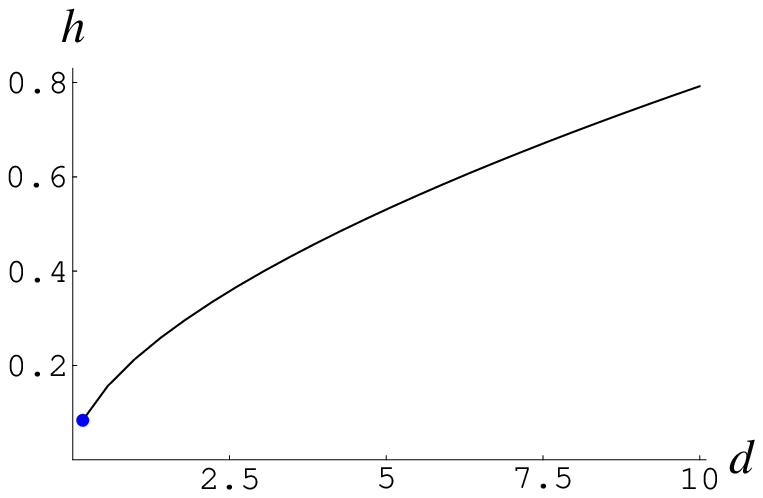,width=7cm} &
\epsfig{file= 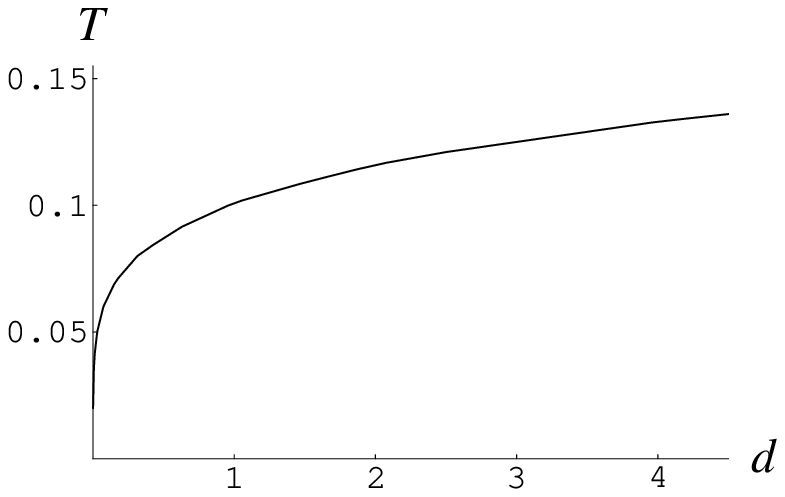,width=7cm}
\end{tabular}
\caption{(a) The phase diagram in the $d$-$h$ plane at $T=0.07$, with the critical point at $d=0.16$ shown as a blue dot, and (b) the critical line in the $T$-$d$ plane.}
\label{d-T_critical_line_fig}
\end{center}
\end{figure}

To further describe the phase transition and to try to understand it better, we look at the value of the magnetic field at which the transition occurs as a function of the density and at fixed temperature.  The results are plotted in Fig.~\ref{d-T_critical_line_fig}(a) and closely fit
\be
\label{phasetransition}
h \sim d^{2/3} \ .
\ee
We also find that even at non-zero temperature the free energy of the phase at large $h$ is well-approximated by
\be
F \sim \frac{d^{2}}{6h} \ .
\ee
As was expected, the behavior here is similar to the zero-temperature case but with finite temperature corrections.  One qualitative difference is in the location of the charges in the high $h$ phase.   While almost all the charges are in the form of smeared D4-branes, they are no longer all near the boundary at $u=\infty$ but are instead spread across all values of $u$.

In terms of the behavior as a function of the temperature, we find that for a given charge density $d$ above a certain temperature there will be no phase transition for any $h$; thus, the line of first-order phase transitions ends at a critical point above which there is a smooth cross-over. The critical line in the $T-d$ plane above which there is no phase transition is given in Fig.~\ref{d-T_critical_line_fig}(b) and behaves at large $d$ as 
\be
T \sim \log d \ .
\ee

It appears that no matter how high the temperature is, one can find a large enough density for which a phase transition will occur at some strong magnetic field.  Of course, at these parameters, for a given $L$, the preferred phase may be the chiral-broken phase.  However, as explained before, we can always take $L$ to be large enough that the chiral-symmetric phase is preferred.

\subsection{Transport properties}
 
The conductivity of this system was computed in \cite{ll1} with the result
\bea
\sigma_{longitudinal} &=& \frac{\sqrt{u_{T}^{8}+h^2u_{T}^{5}+u_{T}^{3}(d-d^{*})^{2}}}{u_{T}^{3}+h^2} \nonumber\\
\sigma_{Hall} &=& -\frac{h(d-d_{*})}{{u_{T}^{3}+h^2}}-\frac{d^{*}}{h} 
\eea
where, as before, $d_{*}=-3ha_{1}(u_{T})$ is the charge carried by the smeared D4 branes.  These charges, located above the horizon in $SU(N_c)$ singlet states, behave as if they do not interact with the dissipative bath. 

The phase transition discussed above is a transition in which $d_{*}$ changes discontinuously.  Consequently, the transport properties also change discontinuously; the Hall conductivity grows and the longitudinal conductivity shrinks.  Note that at zero temperature, in accordance with Lorentz invariance, at any non-zero $h$ the longitudinal conductivity is zero and the Hall conductivity is $-d/h$.  Only at non-zero temperature do the different interactions with the dissipative bath result in a discontinuous conductivity.
  
Because the D4-branes are essentially smeared baryons made of a delocalized collection of quarks, one might have thought that gluons would at some level be able to interact with the individual quarks, leading to dissipation.  However, the thermal bath is holographically associated with the horizon, and since the smeared D4-branes are located above the horizon, their size in the boundary theory is much smaller than the wavelength of the thermal gluons.  As a result, as seen by the gluons, the D4-branes are effectively localized $SU(N_c)$ singlets with which they don't interact.

\section{Some interpretation}

To try to understand some of the physics of this transition, we explore the phase that is dominant at large $h$.  At least some of its properties are similar to the properties of a free, massless fermion gas in four dimensions in its lowest Landau level. 

A massless fermion in four dimensions with charge $e$ in a magnetic field $B$ has the following energy spectrum:
\be
E=\sqrt{k_{1}^{2}+2eBn}
\ee
where $k_{1}$ is the momentum in the direction of the magnetic field and $n$ is a natural number labeling the Landau levels in the transverse directions. The degeneracy of each state is roughly $B$.  If all fermions are in the lowest Landau level, the energy of the gas is given by
\be
E_{tot} \sim B\int_{0}^{k_{max}} dk \ k =\frac{1}{2} B k_{max}^{2} \ ,
\ee
and the density of fermions is 
\be
n_{f}\sim B k_{max} \ ,
\ee
giving a chemical potential and  energy 
\be
\label{freefermionmuandE}
\mu \sim \frac{n_f}{B}, \ \ \ E\sim \frac{n_{f}^{2}}{B} \ .
\ee
The lowest Landau level is not spin degenerate, thus when there's a jump to the lowest Landau level, there is an increase in the magnetization.  We can estimate when the system crosses from the lowest to the first Landau level; this occurs when it is energetically favorable to increase $n$ rather than add to the parallel momentum, meaning $k^{2}_{max} \sim B$, which corresponds to
\be
\label{freefermionphasetransion}
B \sim n_{f}^{2/3} \ .
\ee

Many of the properties of the large $h$ phase agree with those expected from free fermions in the lowest Landau level in four dimensions.  The behavior of the free energy and chemical potential (\ref{muandF}) match the free fermion result (\ref{freefermionmuandE}), and phase transition (\ref{phasetransition}) resembles that of a transition into the lowest Landau level (\ref{freefermionphasetransion}).

On the supergravity side, in the large $h$ phase almost all the baryonic charge\footnote{At $T=0$ this will be all the baryonic charge.} is represented by smeared D4-branes that crept up the D8-brane.  Recently, there have been other indications that smeared D4-branes seem to be related to fermions in the lowest Landau level.  For example, in \cite{sonthom} it was argued that in the confined, large $h$, pion-gradient phase the smeared D4-branes also behave as if they are free fermions and reproduced the thermodynamics of the lowest Landau level.  

As shown in \cite{bll3}, in the chiral-symmetric phase a baryon chemical potential in a background magnetic field generates an axial current.  This anomaly-driven current is also present in a gas of free fermions and is due only to the fermions in the lowest Landau level \cite{metzit}. In the supergravity description, since there is no constant of integration in equation (\ref{EOMa1dec}), the five-dimensional current is located above the horizon; consequently, the axial current is due to the smeared D4-branes rather than the charges at the horizon. 

Furthermore, the smeared D4-branes holographically correspond to bound states of $N_c$ fermions, making up what look like smeared  baryons. This phenomena may be related to the fact that in the lowest Landau level fermions are confined to move only in one spatial dimension, and thus any attractive potential will make them bind, even one which would not bind them in three spatial dimensions.  Such an occurrence has been demonstrated to happen for a different situation related to the magnetically-induced chiral-symmetry breaking \cite{kabat}. 

Because the Sakai-Sugimoto model provides a robust holographic model of QCD, it would be interesting to see if the phenomena described here could be found, for example, at RHIC in a strong magnetic field.  We see the onset of anomaly-driven behavior as soon as a magnetic field is turned on, but this may be an artifact of exactly massless quarks.  In real QCD, a very strong magnetic field may be required before any of these phenomena become manifest.

There are a few non-relativistic fermionic systems which resemble in some way the relativistic system discussed here.  In particular, several condensed matter systems also exhibit first-order quantum phase transitions with discontinuous magnetizations.  In the case of metamagnetism \cite{meta}, the magnetization, which comes from spin ordering, jumps due to a large increase in the density of states near the Fermi surface.  This effect sometimes occurs in paramagnetic materials with itinerant electrons in their outer shells.  The magnetization, as a function of the external magnetic field, can have a large rise and sometimes a discontinuity due to a rapid transition of many fermion spins from anti-parallel to the magnetic field to parallel. 

Another phenomenon that can cause a jump in the magnetization is based on the de Haas-van Alphen (dHvA) effect which is due to orbital angular momentum.  At low temperature free fermions in four dimensions develop oscillations in their magnetization and chemical potential, among other quantities, as a function of the inverse magnetic field.\footnote{For a recent computation of these oscillations for massless fermions see \cite{cme}.}  These oscillations occur when the the Fermi surface sweeps through the Landau levels.  

Because the charges feel both the background magnetic field and field produced by the other charges, the magnetization $M$ of the material is a function of the magnetic inductance $B$, which itself is a function of both $H$ and $M$.  The function $M(B)=M(H,M)$ may therefore have more then one solution for $M$ for a given $H$ if the dHvA oscillations become large \cite{pippard, condon}.  In this case, a first-order phase transition may occur when the system transitions from one Landau level to the next.  This phase transition, which happens in real metals only at very low temperatures, has been observed mainly in diamagnetic metals and is therefore called a diamagnetic phase transition.\footnote{For a review see \cite{gordon}.}  Depending on the geometry of the sample, the phase transition may lead either to a homogenous phase or to an inhomogeneous phase, with multiple Condon domains of different magnetisms. 

The phase transition we find in the SS model appears to have some resemblance to metamagnetism but also some differences.  We start with a paramagnetic state and see a jump in the magnetization at a particular value of the magnetic field.  However, after the jump, the magnetization decreases, while in materials exhibiting metamagnetism, the phase after the transition is ferromagnetic.  In addition, while we find evidence the large magnetic field phase in the SS model is associated with the lowest Landau level, metamagnetism, which occurs at more modest magnetic fields, is not related to lowest Landau level physics.

The diamagnetic phase transition is associated with transitions between Landau levels, and it occurs each time the Fermi surface crosses a Landau level;  we, however, see only one phase transition rather than a whole series.  Indeed, in our system we do not observe the dHvA oscillations at all, even at zero temperature, which may indicate that there is no well-defined Fermi surface.\footnote{The absence of a signal of the Fermi surface was also observed in \cite{parna1}.} Furthermore, in the SS model, we have only a non-dynamical background electromagnetic field, thus $B=H$.  Hence, this is quite different than the diamagnetic phase transition.

We are left then without a truly compelling interpretation of our results.  In addition to serving as a holographic model of QCD, the SS model describes a system of strongly-coupled relativistic fermions, which to some degree reproduces phenomena seen in some condensed matter systems.  There seems to be some analogy with metamagnetism, but this analogy is incomplete.

\centerline{\bf{Acknowledgments}}
We wish to thank Alex Gordon and Ady Stern for very useful explanations and discussions. This work is supported in part by Israel Science Foundation under grant no. 568/05.

\end{document}